# Orientation-Multiplexed Shadow XMCD-PEEM Tomography of 3D Nanomagnets

*Le Zhao*, Jakub Jurczyk, Alberto Anadón, Michael Walser, Michael Foerster, Sandra Ruiz-Gómez, Paweł Nita, Marcin Szpytma, Anna Mandziak, and Amalio Fernández-Pacheco**

L. Zhao, J. Jurczyk, A. Anadón, M. Walser, A. Fernández-Pacheco
Institute of Applied Physics, TU Wien, Wiedner Hauptstraße 8–10, 1040 Vienna, Austria
E-mail: le.zhao@tuwien.ac.at; amalio.fernandez-pacheco@tuwien.ac.at

A. Anadón
Instituto de Nanociencia y Materiales de Aragón (INMA), CSIC-Universidad de Zaragoza, 50009 Zaragoza, Spain

M. Foerster, S. Ruiz-Gómez
ALBA Synchrotron Light Facility, Carrer de la Llum 2–26, 08290 Cerdanyola del Vallès, Barcelona, Spain

M. Szpytma, A. Mandziak
National Synchrotron Radiation Centre SOLARIS, Jagiellonian University, Czerwone Maki 98, 30-392 Kraków, Poland

P. Nita
Faculty of Physics, Astronomy and Applied Computer Science, Jagiellonian University, prof. Stanisława Łojasiewicza 11, 30-348 Kraków, Poland

Funding: European Research Council (ERC), Horizon 2020 (101001290, 3DNANOMAG); Austrian Science Fund (FWF), 10.55776/PIN1629824 (TOMOCHIRAL); TU Wien Blue Sky "Phoenix" program; European Union, Horizon Europe, Marie Skłodowska-Curie Actions (GAP-101108063, 3D-Sky); Agencia Estatal de Investigación (AEI), projects PID2024-157112OB-C55 and PID2024-155385NA-C32.

Imaging magnetization in curved three-dimensional (3D) nanostructures is important for spintronic devices such as non-planar racetracks. Shadow X-ray magnetic circular dichroism photoemission electron microscopy (XMCD-PEEM) can be used to study such geometrically complex structures by probing the photoemission from the shadowed area. However, only the magnetization component parallel to the X-ray beam is detected. Reconstructing the 3D magnetization vector field therefore requires tomographic methods, usually by mechanically rotating the specimen and recording shadow projections at multiple viewing angles. Here we introduce orientation-multiplexed shadow XMCD-PEEM tomography, where nominally identical nanostructures fabricated at different azimuthal orientations provide complementary projections for the reconstruction under a fixed X-ray incidence geometry. The method is applied to reconstruct the 3D magnetization of a series of racetrack-like nanostructures from two shadow projections per state, after initialization along in-plane field directions. The reconstructed fields reproduce the measured contrast and reveal segments whose magnetization changes with the initialization field alongside segments that retain their direction across the four states. For some field directions, the local slope couples in-plane and out-of-plane changes in magnetization. By defining the projection angles during fabrication, this approach enables vector magnetic imaging of curved 3D devices when instrument geometry or electrical contacts restrict sample rotation.

## 1. Introduction

Racetrack memory stores information in magnetic domains that are shifted along nanowires, and its vertical implementation promises storage densities beyond those of planar devices.[1,2] Realizing such architectures requires magnetic tracks that leave the substrate plane and guide domain walls (DWs) through curved and inclined segments.[3–5] More broadly, research in three-dimensional (3D) nanomagnetism has shown that curvature, torsion and spatially varying film thickness can produce effective anisotropies and chiral interactions, giving rise to magnetic textures and DW dynamics that are absent in planar films, as demonstrated in 3D artificial spin ices, helical interconnectors and twisted nanostructures.[6–17]

Understanding and engineering these systems often requires imaging the magnetization of curved 3D surfaces, for which a range of techniques has been developed. In the laboratory, vector field electron tomography in the transmission electron microscope, based on electron holography or Lorentz microscopy, has resolved the 3D magnetic induction of nanostructures,[18–21] and dark-field magneto-optical Kerr magnetometry and extended depth-of-field Kerr microscopy are being adapted to non-planar samples.[22,23] At synchrotron facilities, transmission X-ray vector tomography and laminography have revealed complex 3D magnetic configurations,[24–28] and full-field soft X-ray microscopy, including shadow X-ray magnetic circular dichroism (XMCD) photoemission electron microscopy (PEEM), has been combined with sample rotation to retrieve the magnetization of curved thin films.[29] Although XMCD-PEEM images of planar films recorded at several sample azimuthal orientations have been combined to reconstruct vector maps of the magnetization,[30] shadow XMCD-PEEM studies of 3D nanostructures have so far mostly relied on imaging from a single viewing direction.[14,31–34] At grazing X-ray incidence, the structure casts a shadow on the substrate, and the photoelectrons emitted from the shadowed region encode the transmission of circularly polarized X-rays through the magnetic material. Owing to XMCD, the shadow contrast is sensitive to the magnetization component along the beam, and the grazing geometry stretches the projection along the beam direction, which makes the technique particularly attractive for 3D nanostructures. Combined with sample holders for *in situ* magnetic fields,[35] shadow XMCD-PEEM has enabled the observation of DW automotion and pinning in 3D helical interconnectors[14] and of complex magnetic textures in 3D nanoarchitectures.[34] As a single projection generally does not provide enough information to recover the 3D magnetization configuration directly, such measurements often rely on comparison with micromagnetic simulations, which limits the interpretation of novel magnetization states.

Quantitative reconstruction of the magnetization field with more than one unknown component has two main requirements. First, projections from two or more independent viewing directions are needed. In conventional tomography, these are obtained by mechanically rotating the specimen,[29] which requires the magnetic state to remain unchanged during rotation, or to be reproduced at every angle. The structure also needs to be relocated and refocused after each rotation. Second, the reconstruction must identify which elements of the 3D magnetic surface contribute to each pixel in the shadow image. This requires an indexable 3D model of the actual sample geometry, which may differ from the nominal design.

Here, we address both requirements with orientation-multiplexed shadow XMCD-PEEM tomography. Instead of rotating a single specimen, we fabricate a series of nominally identical but differently oriented nanostructures by focused electron beam induced deposition (FEBID)[36–41]. These replicas provide complementary projections under a fixed X-ray incidence geometry. The actual geometry of each structure is reconstructed from multi-view scanning electron microscopy (SEM) silhouettes, and a comparison of five replicas shows that they agree closely, especially at the top surface that carries the magnetic film. Combining shadow XMCD-PEEM images of two replicas oriented 45° apart with a geometry-constrained forward model, we reconstruct the surface magnetization of Ta/CoFeB/Ta-coated racetrack-like nanostructures after applying in-plane fields along four orthogonal directions. The reconstructions reproduce the measured shadow contrast and reveal how the device responds to the field, including segments whose magnetization switches with the field and segments whose magnetization state is almost the same across all field initialization protocols. The technique presented here broadens the scope of shadow tomography to complex 3D nanomagnets. Defining the projection angles during fabrication can reduce the time spent repositioning the sample, give access to viewing angles beyond the rotation range of the stage, and allow electrical probes to remain in contact during imaging. We examine how closely the fabricated replicas match and discuss the magnetic initialization needed to obtain equivalent states in their respective local frames.

## 2. Results and Discussion

### 2.1. Orientation Multiplexing of Nominally Identical Nanostructures

Here we compare two ways of acquiring angular information in **Figure 1**. In projection-angle scanning (conventional approach, **Figure 1a**), a single device is rotated on the stage, so that the incident beam probes the structure from different azimuthal directions, and the shadow projection changes accordingly. In device-orientation multiplexing (this work, **Figure 1b**), nominally identical devices are fabricated with different azimuthal orientations and imaged at fixed X-ray incidence, so that each device contributes a different projection without any mechanical stage rotation. **Figure 1c** shows the five devices, D1–D5, studied in this work. They are written by FEBID at different azimuthal angles and include mirror-related configurations (**Table S1**, Supporting Information). The devices are located close to each other on the substrate, so that they can be imaged simultaneously within the same PEEM field of view. The two devices used for magnetic reconstruction, D1 and D2, have the same handedness, with D2 oriented 45° clockwise with respect to D1.

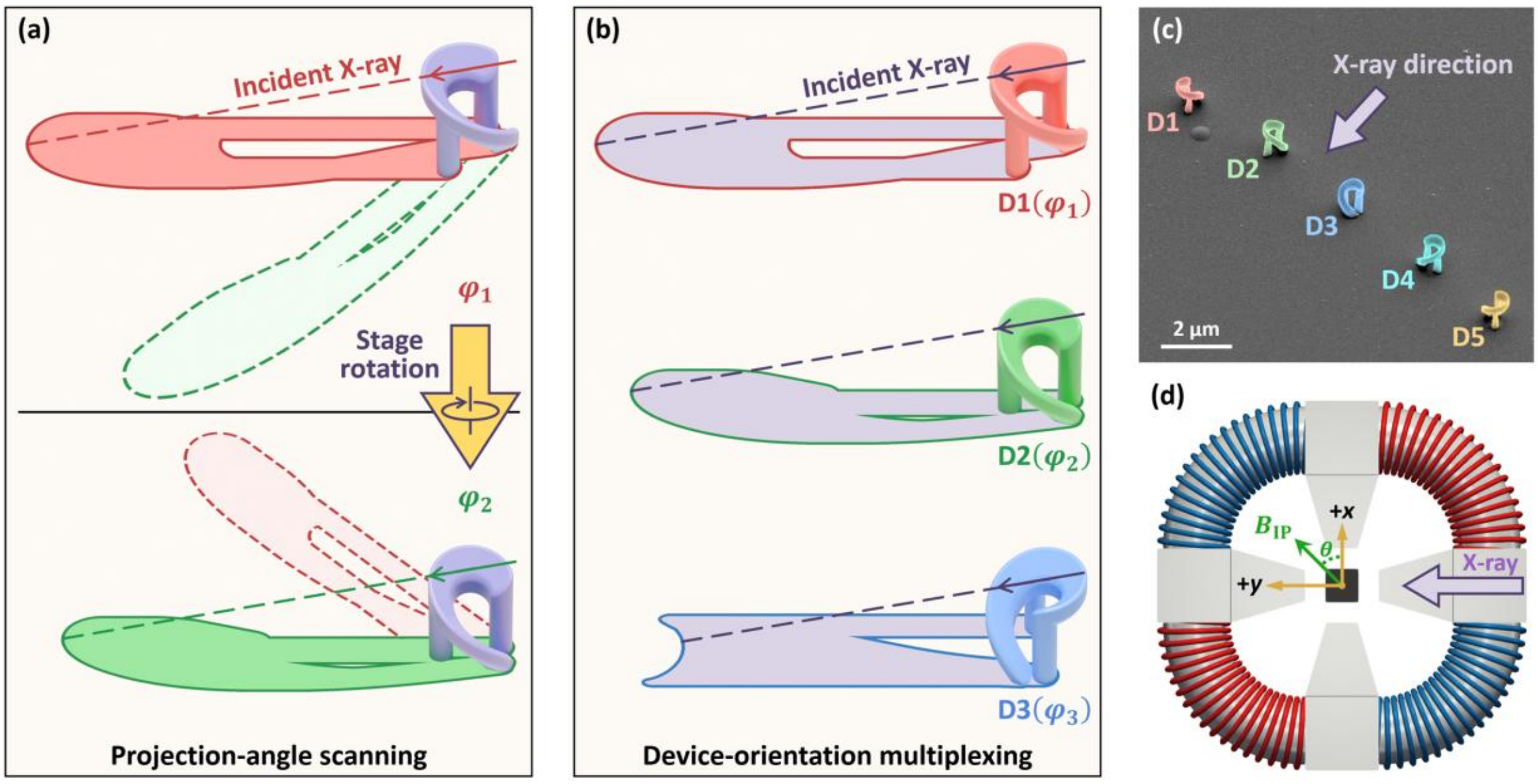

**Figure 1.** Orientation multiplexing for shadow magnetic imaging. a) Projection-angle scanning (conventional approach). Rotating a single three-dimensional device changes its azimuth relative to the incident X-ray beam and provides different shadow projections. Red and green indicate two of these azimuthal imaging projections, $\varphi_1$ and $\varphi_2$. Solid and dashed outlines of the same color represent the same shadow projection related by the indicated stage rotation. b) Orientation multiplexing (approach used here). Nominally identical devices fabricated at different azimuthal angles provide complementary projections at fixed X-ray incidence. Devices D1–D3 illustrate three relative orientations, $\varphi_1$–$\varphi_3$. This approach provides the angular information needed for shadow XMCD-PEEM tomography at fixed stage orientation. It therefore avoids the mechanical stage rotation needed to acquire the different projection views in the conventional approach. c) False-colored scanning electron microscopy (SEM) image of five devices, D1–D5, with different azimuthal orientations and mirror-related configurations on the substrate. The arrow indicates the incident X-ray direction used

for magnetic imaging. d) Schematic of the quadrupole electromagnet surrounding the central sample position, used to prepare states with arbitrary in-plane field directions.

Magnetic states are prepared with a quadrupole electromagnet surrounding the sample (**Figure 1d**), which generates in-plane fields along arbitrary directions. In the multiplexed approach, rotating the field electrically takes over the role of rotating the sample. Because D1 is rotated by 45° counterclockwise with respect to D2, the state of D2 after a field applied along a given in-plane direction is equivalent to the state of D1 after the same field rotated counterclockwise by 45°. A series of fields applied in 45° steps provides, for every field direction relative to the devices, shadow projections of equivalent states from two different azimuthal angles. In this way, the invariance of a single magnetic state under mechanical rotation is replaced by two requirements that can be addressed independently: geometric equivalence of the replicas, which is established during fabrication, and equivalent magnetic states, which can be prepared by matching the field direction relative to each replica before imaging.

Orientation multiplexing thus moves the angular sampling from the measurement to the fabrication step. As the orientations are predefined by the FEBID writing pattern and remain fixed during acquisition, projections from several azimuthal angles can be recorded in the same field of view, without relocating, refocusing or readjusting the magnification after each rotation. This may shorten the acquisition time and reduce the cumulative X-ray exposure of each structure, which is relevant for beam-sensitive nanostructures. Keeping the stage fixed is also advantageous when rotation would disrupt the contact between the device and electrical probes during imaging.[42] Replicas can also be pre-oriented at angles that a conventional sample stage cannot easily reach, including tilts with respect to the substrate normal. Orientation multiplexing can also be combined with stage rotation, providing several additional projections per rotation step. These advantages come at the price of the two equivalence requirements above, together with a set of orientations that is fixed once the sample is fabricated.

To assess the feasibility of this approach, we first establish the actual 3D geometry of the devices and quantify their similarity.

### 2.2. Three-Dimensional Geometry from Multi-View SEM Silhouettes

Each investigated device is a 3D racetrack-like nanostructure, in which a ribbon-shaped track with a width of approximately 200 nm rises from the substrate along a 60° slope and continues into a semicircular section, whose surface is additionally tilted inward by 45° in the radial direction. The elevated part is supported by two vertical pillars, and the total height is approximately 900 nm. The geometry of each device is reconstructed from SEM images acquired at 24 azimuthal angles from 0° to 345° in 15° steps (**Figure 2a**). After a coarse alignment based on geometric features, each image is converted into a binary silhouette that separates the projected device from the background and from internal openings (**Figure 2b**). Since every silhouette confines the device to the volume obtained by back-projecting it along the viewing direction, the intersection of the back-projected silhouettes defines the admissible 3D volume (**Figure 2c**). This method is known as "shape from silhouette".[43] The reconstruction is refined iteratively by removing voxels whose projections are inconsistent with the measured silhouettes and by aligning each image to the reprojection of the current volume, which compensates for small displacements of the rotation center between acquisitions (see **Experimental Section**).

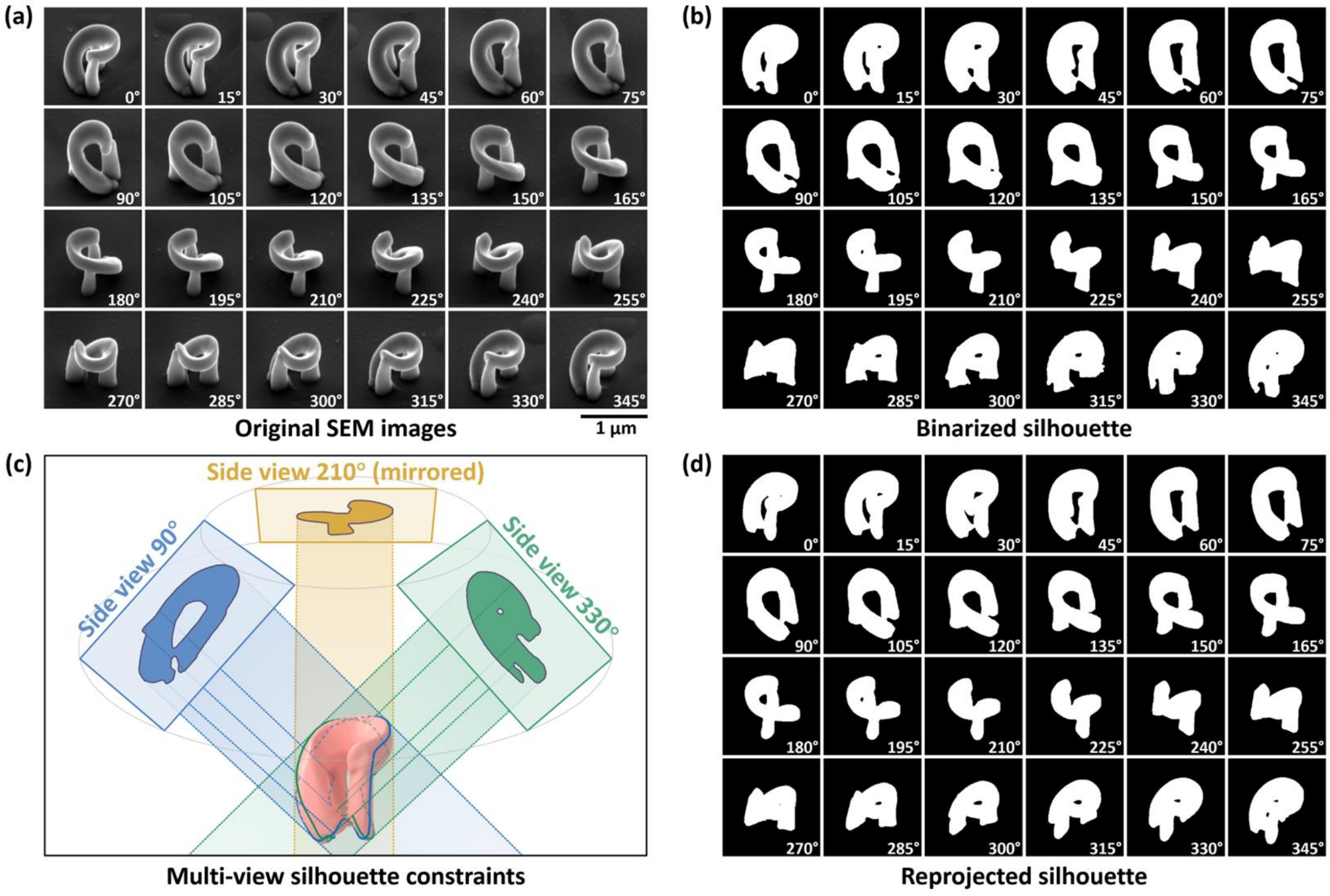

**Figure 2.** Three-dimensional geometry constrained by multi-view SEM silhouettes. a) SEM images of a representative device acquired at 24 azimuthal angles from 0° to 345° in 15° steps. b) Corresponding binary silhouettes, with white denoting the projected device and black the background and internal openings. c) Illustration of the geometric constraints supplied by multiple views. Back-projected silhouettes restrict the admissible three-dimensional volume. The blue, yellow and green projection planes, together with their associated rays and surface contours, illustrate views at 90°,

210° and 330°, respectively. The 210° view is mirrored for display, as labeled. d) Silhouettes obtained by reprojecting the reconstructed geometry at the same angles as in (a,b), allowing comparison with the measured outlines and openings. Panels (a), (b) and (d) share the same scale bar.

To assess the reconstruction, we reproject the reconstructed geometry at the 24 acquisition angles (**Figure 2d**). The reprojected silhouettes reproduce the measured outlines well, including the internal openings, the track itself and its supports (**Figures S1–S5**, Supporting Information). As the silhouettes constrain the outer envelope of the structure, the reconstruction is most reliable for the outer or convex surfaces, which in our case are also the surfaces that receive the sputtered magnetic film. The reconstructed volume thus provides the indexable 3D model required to map each surface facet onto the detector pixels in the XMCD-PEEM measurements.

## 2.3. Geometric Equivalence of the Replicas

The geometries of D1–D5 reconstructed from the SEM images are shown in **Figure 3a** in their original relative orientations. For a quantitative comparison, each device is rotated about the vertical axis, mirrored where required and translated into a common frame (D1*–D5*, **Figure 3b**). An averaged geometry is then obtained from the mean of the signed-distance fields of the five aligned devices (**Figure 3c**).

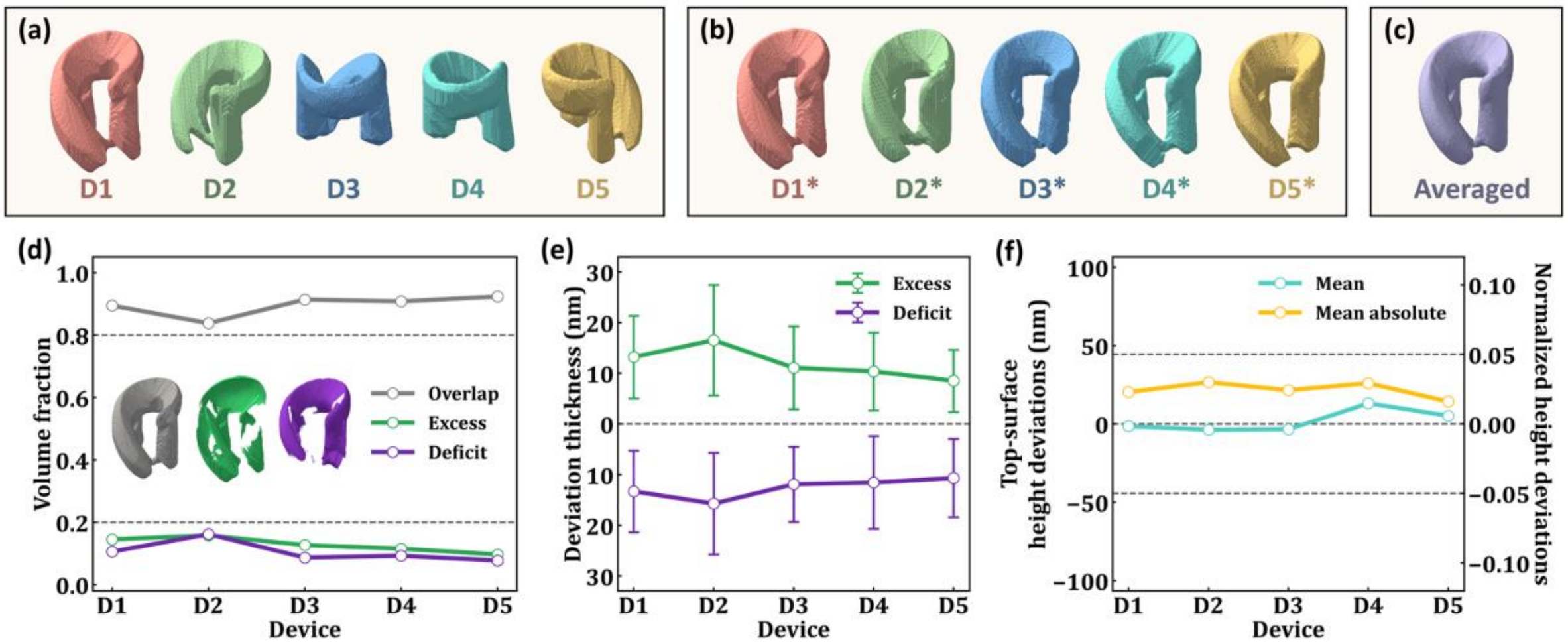

**Figure 3.** Geometric comparison of the orientation-multiplexed devices. a) Three-dimensional geometries reconstructed for D1–D5, shown in their original relative azimuthal orientations and mirror-related configurations. Device colors follow **Figure 1c**. b) The same geometries after rotation, mirror reflection where required, and positional alignment into a common comparison frame, denoted D1*–D5*. c) Averaged geometry obtained from the mean of the signed-distance fields of the five

aligned devices. d) Overlap, excess and deficit volume fractions for each aligned device compared with the averaged geometry. All three quantities are normalized by the volume of the averaged geometry. Insets show D1 as a representative example, with these regions in gray, green and purple, respectively. e) Mean geometric deviation thicknesses for the excess and deficit regions, plotted above and below zero, respectively. f) Mean and mean absolute height deviations from the averaged geometry within the selected top-surface projection region. Top surfaces are extracted using a $75^\circ$ inclination limit relative to the $xy$ plane. The right axis normalizes these quantities by the full height of the averaged geometry.

The volume comparison in **Figure 3d** shows that every replica shares more than 80% of the volume of the averaged geometry. The excess and deficit regions correspond to thin layers distributed over the surface rather than missing or additional structural elements, as shown for device D1 in the inset. The mean thicknesses are only of the order of 10 nm for all devices (**Figure 3e**). In particular, the top surface, which receives most of the magnetic film in magnetron sputtering at normal incidence and defines the outline of the shadow at grazing incidence, is also closely reproduced. The mean top-surface height of every replica differs from that of the averaged geometry by less than 14 nm, about 1.5% of the total device height, and the mean absolute height deviation is at most 30 nm (**Figure 3f**). These differences may arise from the anisotropic precursor flux from a gas injection system, which makes FEBID growth sensitive to the orientation of each device relative to the injector.[44,45] The agreement within a few tens of nanometers and without systematic height offsets across the replicas shows that the device geometry is reproducible under the writing conditions used here (**Experimental Section**). It therefore supports the use of an averaged geometry as a common structural model. For the magnetic reconstruction of D1 and D2, we use the geometry averaged over these two devices only, obtained in the same way.

### 2.4. Shadow XMCD Imaging and Geometry-Constrained Reconstruction

The shadow XMCD-PEEM images in **Figure 4a,b** were recorded for D1 and D2, respectively, at remanence after in-plane fields of approximately 30 mT were applied for 1 s along four orthogonal directions (see **Experimental Section**). The structural image $I^+$ (center) shows the elongated shadow cast by the structure at a grazing incidence of 16°, and the dashed contour marks the shadow of the geometry averaged over D1 and D2, projected along the same beam direction and registered to the measured image. The colored magnetic images display the XMCD contrast

$$C = \frac{I^+ - I^-}{I^+ + I^-} \quad (1)$$

where $I^+$ and $I^-$ are the intensities recorded with opposite X-ray helicities. We take D2 as the reference device, and field directions and axes refer to its frame, in which the in-plane projection of the incident beam is parallel to $y$. For D2, the shadow contrast changes markedly between $+B_y$ and $-B_y$, whereas $+B_x$ and $-B_x$ produce a similar contrast, as expected for a projection that is insensitive to $M_x$. As D1 is rotated by 45° with respect to D2, magnetic fields rotated counterclockwise by 45° prepare the equivalent states in D1 (**Section 2.1**). These field initializations and the corresponding D1 frame are denoted by starred axes. In the D1 frame, the in-plane beam direction differs by 45° from that in the D2 frame, and the contrast in the main part of the shadow reverses between $+B_{x^*}$ and $-B_{x^*}$. However, the region near the tip of the shadow, which corresponds to the uppermost part of the track, keeps the same contrast sign for all four initializations in both devices, indicating that the 30 mT initialization field is not strong enough to change the remanent state of this part of the structure. For each state, the pair of images provides two complementary projections of the magnetization, and changes that are hidden in one projection can be observed in the other.

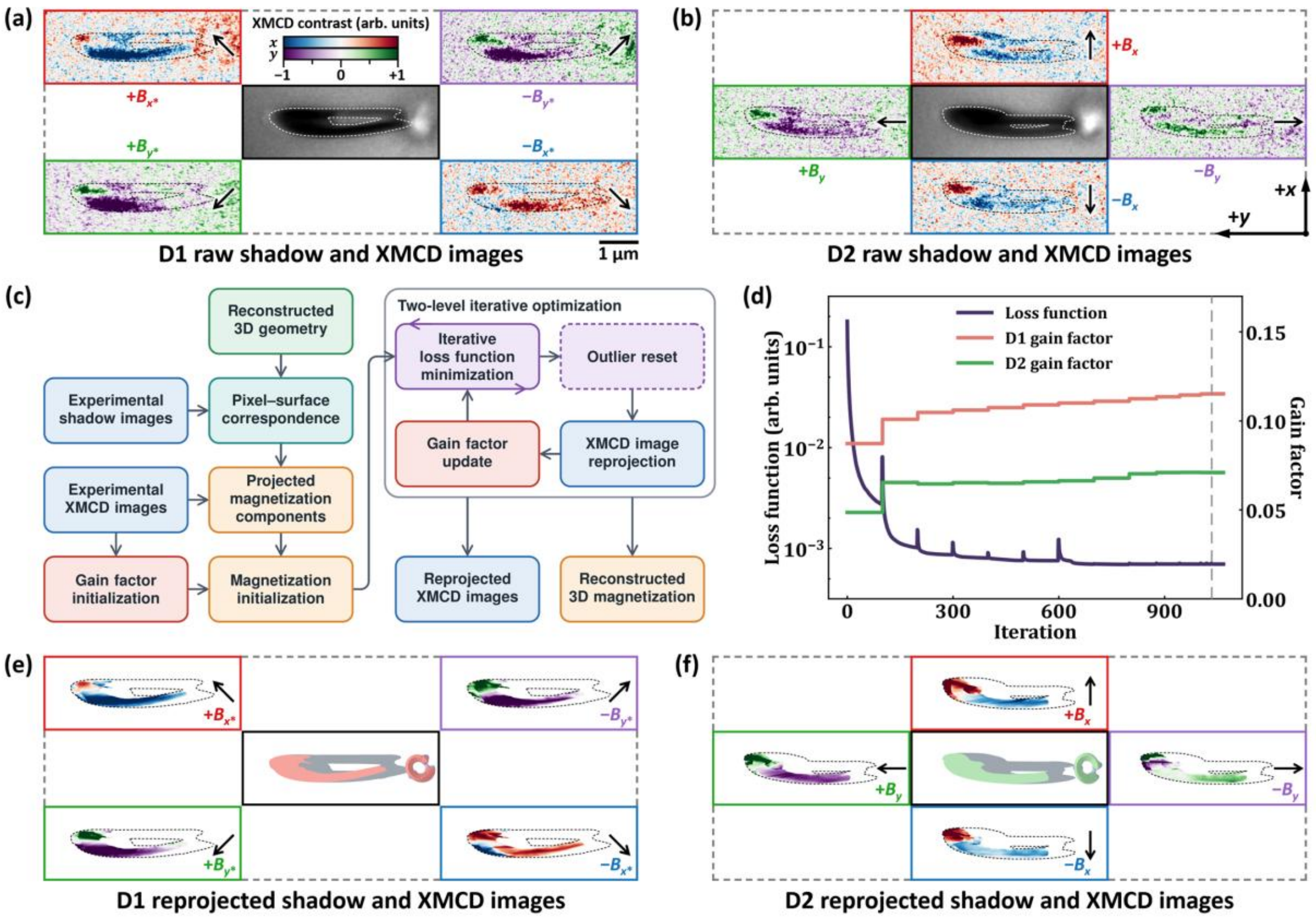

**Figure 4.** Shadow XMCD-PEEM measurements, iterative reconstruction and forward projections. a,b) Measured shadow XMCD-PEEM images of D1 and D2, respectively, at remanence after an in-plane initialization field along the directions indicated by black arrows, arranged around a representative structural shadow image, $I^{+}$. Frames of the same color mark equivalent magnetic states. The red–blue and green–purple palettes show the XMCD contrast (Equation 1) for the $x$- and $y$-initialization series, respectively. Dashed contours mark the structural projection. Axes refer to the D2 frame, and starred axes to the corresponding D1 frame. c) Geometry-constrained reconstruction workflow. The registered three-dimensional geometry links image pixels to surface facets, the measured XMCD images are back-projected onto these facets, and a two-level iterative optimization determines the gain factors and the three-dimensional magnetization vector field. d) Example convergence for the remanent state following a magnetic field applied along $-B_y$ (**Figure 5b**, left). The loss function is shown in purple on the logarithmic left axis, and the D1 and D2 gain factors on the linear right axis. The horizontal coordinate counts the overall iterations. The vertical dashed line marks the transition from the first to the second phase of iterative optimization, after which the outlier reset during the outer iteration is disabled. e,f) Forward-projected XMCD contrast for D1 and D2 from the reconstructed fields, arranged to match (a,b). Central schematics highlight the magnetic surface and its projected coverage considered during reconstruction, with the remaining structural projection in gray. Panels (a), (b), (e) and (f) share the same scale bar and the same two-color bars.

The reconstruction follows the workflow shown in **Figure 4c**. The geometry averaged over D1 and D2 is registered to the measured structural shadow of each device by a two-dimensional similarity transformation, consisting of a translation, an in-plane rotation and an isotropic scaling. This establishes the correspondence between every detector pixel and the surface facets crossed by the X-rays that reach the substrate at that pixel (see **Section S4 and Table S2**, Supporting Information). The magnetic surface is restricted to the portion of the continuous upper surface of the track with a local inclination below 75°, as this surface receives most of the directionally deposited film (see **Experimental Section**).[14,40,46] It is shown in color in the central schematics of **Figure 4e,f**.

Two approximations make the problem tractable with only a pair of projections: the magnetization of each facet $f$ is described by a unit vector $\boldsymbol{m}_f$ lying in the local surface plane,[11,29] and the XMCD contrast is taken to be linear in the projection of the magnetization onto the X-ray propagation direction $\boldsymbol{k}_v$ of view $v$, with contributions from different crossings of the magnetic surface along the beam path adding up.[29,32] The predicted offset-corrected contrast at pixel $p$ is then

$$\hat{y}_{vp} = -g_v \sum_{f=1}^{F} H_{vpf} \, (\boldsymbol{m}_f \cdot \boldsymbol{k}_v) \tag{2}$$

where $F$ is the number of magnetic facets, $H_{vpf}$ is the projected area fraction of facet $f$ within pixel $p$ and $g_v$ is a gain factor that converts the dimensionless projection into the measured contrast of view $v$. Here, $\boldsymbol{m}_f = \boldsymbol{M}_f / M_s$ is the magnetization normalized by the saturation magnetization $M_s$, and $\boldsymbol{k}_v$ is the unit vector along X-ray propagation. The minus sign accounts for the stronger absorption and hence darker transmitted shadow in $I^+$ when magnetization is parallel to the incident beam. Predictions are compared with the offset-corrected measurement $y_{vp} = C_{vp} - b_v$, where $b_v$ is the nonmagnetic background offset of view $v$, and $C_{vp}$ is the measured XMCD contrast at that pixel before background subtraction (**Equation 1**).
Using these projection relations, the measured XMCD images are back-projected to estimate the beam-parallel magnetization component of each contributing facet. Initial gain factors are estimated from the strongest measured contrast in each view, and an initial magnetization is obtained from an approximate tangent-plane least-squares estimate of Equation 2, in which the unit-length constraint is relaxed. After this initial estimate, each magnetization vector is constrained to unit length, so that only its direction is refined. A two-level iterative optimization then refines both the gain factors and the magnetization directions. At the inner level, the magnetization is optimized for fixed gain factors by minimizing a loss function that combines the misfit between measured and reprojected contrast ($E_d$) with weak penalties on abrupt changes between neighboring facets ($E_c$) and on local surface magnetic charges ($E_{\mathrm{div}}$). At the outer level, the gain factors are updated by fitting the calculated projections to the measured contrast in each view. During a first phase, the magnetization of any isolated facet whose direction differs by more than 90° from that of its surroundings is reset to the local average (outlier reset). In the second phase, the iteration continues without further reset until the gain factors are stable (see **Section S6**, Supporting Information).
**Figure 4d** shows the convergence for the state prepared along $-B_y$ ($-B_{y^*}$). The loss function decreases by about two orders of magnitude before leveling off, with small transient increases after each gain update, and the gain factors of D1 and D2 converge to 0.115 and 0.071, respectively. The other three states converge according to the same criterion (**Table S3**, Supporting Information). The XMCD contrast forward-projected from the reconstructed magnetization (**Figure 4e,f**) reproduces the main features of the measurements in both devices, including the sign of the contrast in the different parts of the shadow, its changes between opposite initializations, the persistent contrast near the shadow tip, and the positions along the shadow at which the contrast changes sign. The penalty terms $E_c$ and $E_{\mathrm{div}}$ also help to smooth the final reconstruction, so that the forward-projected XMCD images do not reproduce the pixel-level noise in the measured images. The agreement in these main contrast features provides a necessary consistency check of the forward model and the reconstructed field, which can now be examined in three dimensions.

### 2.5. Field-Prepared Magnetization of the 3D Track

The orthographic top views in **Figure 5a,b** show the reconstructed magnetization at remanence after initialization along the four field directions. Over most of the surface, the arrows run along the track, showing that the reconstructed magnetization predominantly follows its long axis. This is consistent with the shape anisotropy of a narrow magnetic strip and provides a physical check on the reconstruction, since this energy term is not explicitly included in our image-based reconstruction. After initialization along $-B_x$ (**Figure 5a**, left), the left part of the loop and the inward-curling end of the track at the bottom show negative $M_x$, whereas a segment along the upper arc retains positive $M_x$. These regions are separated by transition zones in which $M_x$ changes sign. After initialization along $+B_x$ (**Figure 5a**, right), positive $M_x$ extends over most of the surface, except near the lower end of the left part of the loop. The upper arc thus has positive $M_x$ after both initializations, whereas the remaining segments switch with the field. A similar behavior is found after the $y$ initializations (**Figure 5b**), where the left part of the loop changes from weakly negative $M_y$ after $-B_y$ to positive $M_y$ after $+B_y$, whereas the right-hand segment of the loop shows negative $M_y$ in both states.

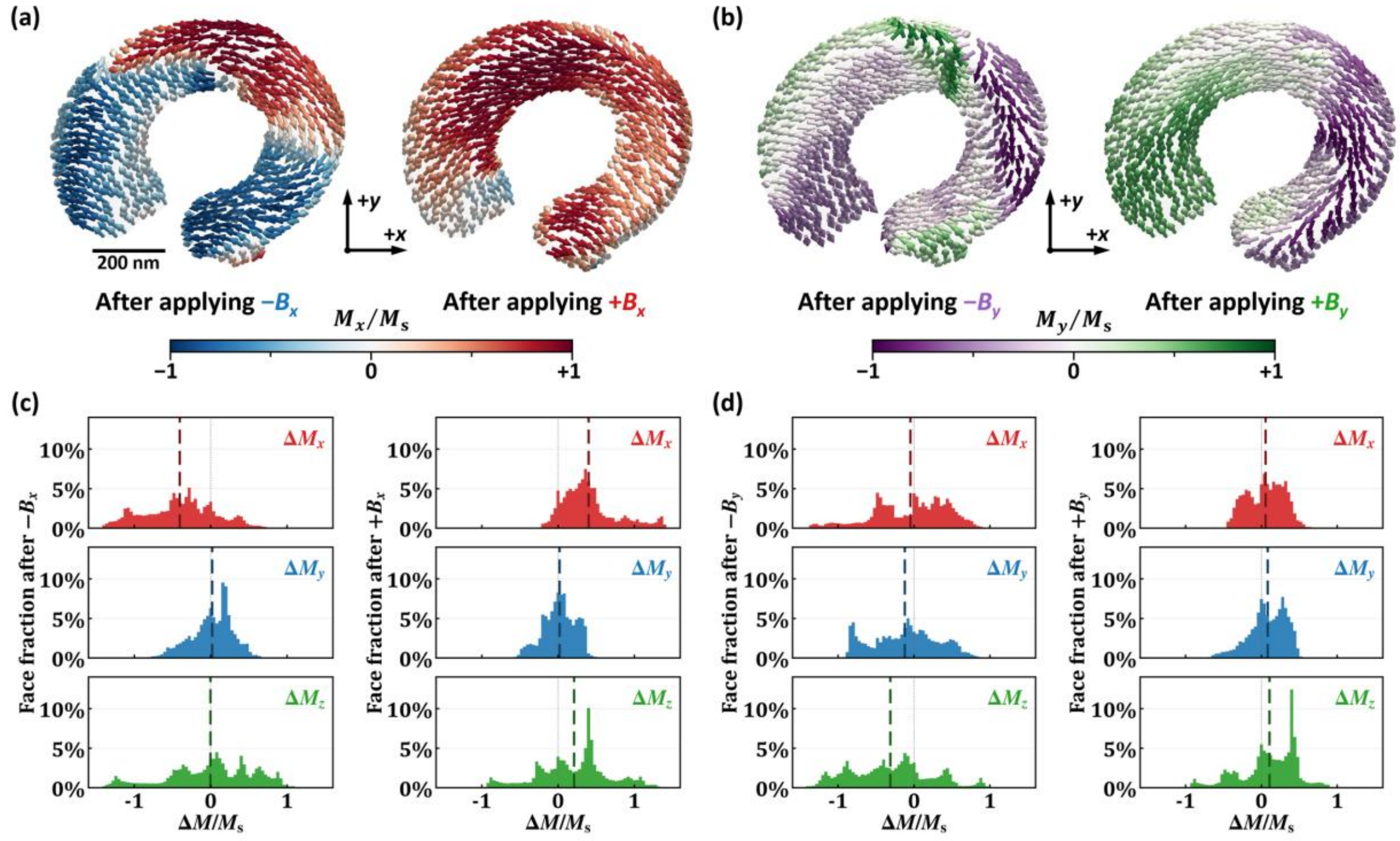

**Figure 5.** Reconstructed surface magnetization and state-dependent component distributions. a) Orthographic top views of the reconstructed magnetization following initialization along $-B_x$ (left) and $+B_x$ (right) in the displayed frame. Arrows show the local magnetization direction and are colored by $M_x/M_s$. b) Corresponding views following initialization along $-B_y$ (left) and $+B_y$ (right), with arrows colored by $M_y/M_s$. c,d) Component distributions for the states shown in (a,b), after subtracting the same four-state averaged reference at every matched surface face, respectively. Colored vertical dashed lines indicate unbinned component means. Panels (a) and (b) share the same scale bar.

To separate the field-dependent part of the response from the component shared by all initializations, we subtract at every facet $f$ the mean of the four reconstructed states, $\bar{\boldsymbol{m}}_f = \frac{1}{4}\sum_{s=1}^{4} \boldsymbol{m}_{s,f}$, and analyze the deviations $\Delta \boldsymbol{M}_{s,f}/M_s = \boldsymbol{m}_{s,f} - \bar{\boldsymbol{m}}_f$, where $s$ labels the initializations $\pm B_x$ and $\pm B_y$ (**Figure 5c,d**). For the $x$ initializations, $\Delta M_x$ is shifted to negative values after $-B_x$ and to positive values after $+B_x$, with mean values of $-0.41$ and $+0.40$, respectively, whereas the mean of $\Delta M_y$ remains close to zero (0.02) in both states. For the $y$ initializations, the mean of $\Delta M_y$ changes from $-0.12$ after $-B_y$ to $+0.08$ after $+B_y$, and the mean of $\Delta M_z$ changes from $-0.31$ to $+0.11$.

The response to the initialization field along $y$ is thus weaker than that to the field along $x$ and is accompanied by a substantial change in $M_z$. This coupling between in-plane and out-of-plane components is consistent with the three-dimensional geometry of the track, which curves out of the substrate plane. As the magnetization is confined to the local surface, reversing it along an inclined segment of the track changes both its in-plane and its vertical component. In the displayed frame, the 60° slope of the track rises approximately in the $y$–$z$ plane. A field along $y$ therefore has a substantial projection along the slope and changes $M_z$ as well, whereas a field along $x$ is nearly perpendicular to the uphill direction of the sloped segment and therefore has little effect on $M_z$.

Several mechanisms may explain why the magnetization in some segments does not follow the initialization field. These include pinning at local variations of geometry or film thickness, such as near the junctions with the supporting pillars,[14,47] and geometry-driven DW motion after the field is removed.[14] Moreover, the in-plane applied field of approximately 30 mT is limited, and on inclined segments only its projection onto the local surface acts on the local tangential magnetization, so that some regions may not be fully saturated. The present static measurements at remanence do not distinguish between these possibilities. The remanent states show that 3D racetracks respond to applied fields in a more complex way than planar devices. Notably, orientation multiplexing makes such configurations accessible from a small number of projections and without sample rotation.

The approach can also be extended to additional directions. Including further replicas, such as D3–D5, can provide additional projections to further constrain the reconstruction. Replicas written with opposite handedness can also give access to geometric mirror images of the same structure within one experiment. Combining orientation multiplexing with a limited number of stage rotations would further increase the angular coverage. Incorporating micromagnetic simulations into the reconstruction could further constrain the solution to physically plausible magnetization states.[48] The main limitation of this approach is the assumption that nominally identical replicas adopt equivalent magnetic states. Fabrication tolerances, which are a few tens of nanometers here, and local pinning sites can cause replicas to adopt slightly different states, particularly when switching is stochastic. Each reconstruction should therefore include a comparison of the replica geometries and a check of the agreement between measured and reprojected contrast. Since nominally identical structures can now be fabricated at different orientations by FEBID and other 3D nanoprinting techniques, the approach can also be transferred to other X-ray and electron imaging methods in which sample rotation is restricted.

## 3. Conclusion

We have introduced orientation-multiplexed shadow XMCD-PEEM tomography, in which nominally identical nanostructures fabricated at different azimuthal orientations provide complementary projections at fixed sample orientation. Multi-view SEM silhouettes provide an indexable model of the actual geometry of FEBID-grown racetrack-like nanostructures, and a comparison of five replicas shows that their top surfaces agree with a common averaged geometry on the scale of a few tens of nanometers. By combining shadow projections from two replicas with a geometry-constrained forward model, we reconstructed the surface magnetization of Ta/CoFeB/Ta-coated tracks after in-plane field application along four directions. The reconstructions reproduce the measured shadow contrast and reveal a spatially varying remanent magnetization along the curved track. After initialization with in-plane fields of approximately 30 mT, some segments align along the initialization field, while others retain the same magnetization direction across all four initializations. For certain in-plane field directions, the remanent magnetization changes in both planar and inclined sections of the track, with the local slope coupling the in-plane and out-of-plane components of the magnetization.

Under the assumption that magnetization is uniform through the film thickness and confined to the local surface plane, the present reconstruction requires projections from at least two differently oriented structures. However, the method is not restricted to this assumption. For example, when magnetization is not confined in-plane and may also contain a component normal to the local surface, resolving the full vector would then require projections from at least three differently oriented structures that provide linearly independent beam directions in a common device frame. Additional projections from suitably separated structural orientations could further improve reconstruction reliability.[30]

More generally, the method could also be used to reconstruct magnetization in bulk systems.[32,49] Unlike magnetic thin films on curved surfaces,[29] these systems require magnetization vectors throughout the volume to be refined, and therefore a larger series of differently oriented structures would be needed. More extensive angular sampling would provide further information for improving spatial resolution and reconstruction accuracy in these cases.[24]

The reconstruction itself could be further simplified by including a planar reference with the same magnetic coating within the field of view during the experiment. Its shadow contrast with a known magnetization direction would provide an experimental calibration of the gain factors, which could then be used directly in Equation 2 to reconstruct the magnetization without the need for an outer gain-fitting loop.

Orientation multiplexing thus uses the design freedom of three-dimensional nanofabrication to tailor the angular information available to the experiment. Combined with reconstruction of the actual device geometry, it provides a practical route to vector magnetic imaging under restricted sample rotation. This capability expands the experimental scope of shadow XMCD-PEEM tomography for three-dimensional spintronic architectures, including vertical racetracks, and creates opportunities to study electrically contacted devices while keeping their measurement configuration intact.

## 4. Experimental Section

*Sample Fabrication*: Three-dimensional racetrack-like scaffolds were fabricated by FEBID[37,50] with the MeCpPt(Me)$_3$ precursor on a Si/SiO$_2$ substrate in a TESCAN CLARA system at an accelerating voltage of 30 kV and a beam current of 30 pA. The structures were written layer by layer from a computer-designed model using the F3AST software (FEBID 3D Algorithm for Stream File generation).[39] Five replicas of identical design were written at different azimuthal orientations (**Figure 1c**). Relative to D1, D2 and D3 are rotated clockwise by 45° and 135°, respectively. D4 is a copy of D3 obtained by reversing the $x$ coordinate in the frame defined in **Figures 4** and **5**, leaving y and z unchanged. D5 is rotated by 45° clockwise relative to D4. A Ta (5 nm)/$Co_{40}Fe_{40}B_{20}$ (50 nm)/Ta (10 nm) stack was subsequently deposited by magnetron sputtering at normal incidence (AJA International sputtering system, base pressure $\approx 10^{-8}$ Torr). The bottom Ta layer serves as a seed layer. The top Ta layer protects the stack and suppresses the direct XMCD signal from the CoFeB film on the substrate, so that the contrast in the shadow arises only from the X-rays transmitted through the 3D structure.[14] The thicknesses are nominal values for a planar substrate. On inclined surfaces of the 3D structure, the local film thickness could be reduced owing to the directionality of the deposition.

*Structural Imaging and Geometry Reconstruction*: SEM images of the devices were acquired after the XMCD-PEEM measurements at a tilt angle of 40° for 24 azimuthal orientations in 15° steps. The images were aligned horizontally and vertically using geometric features of the structures, median-filtered, binarized with the adaptive local threshold of Phansalkar et al.,[51] and refined by constrained region growing. The volume was reconstructed on a voxel grid with a voxel size of 4.88 nm by shape from silhouette.[43] Voxels whose projections were inconsistent with the measured silhouettes were removed iteratively, and the image alignment was refined between iterations by comparison with the reprojected volume. The outer surface was then extracted as a triangulated mesh and smoothed. For the comparison in **Figure 3**, the devices were aligned by rotation about the vertical axis, mirror reflection where required and translation. The averaged geometry was obtained by averaging the signed-distance fields of the five aligned devices on a common grid and extracting the zero-level surface. The deviation thickness was evaluated as the mean local thickness of the excess and deficit regions, with its standard deviation shown as error bars in **Figure 3e**. The top-surface height deviation was evaluated as the height difference between a device and the averaged geometry along vertical rays whose first intersection with each geometry lies on its top surface.

*XMCD-PEEM Measurements*: Shadow XMCD-PEEM measurements were performed with the Elmitec LEEM/PEEM III end station of the DEMETER beamline at the National Synchrotron Radiation Centre SOLARIS.[52] Circularly polarized X-rays were incident at a grazing angle of 16° with respect to the substrate plane. The photon energy was set to 708.5 eV at the Fe $L_3$ absorption edge, selected from X-ray absorption spectra acquired at remanence. Before mounting, the sample was exposed to an out-of-plane field of approximately 50 mT along $+z$ for saturation. In the microscope, in-plane magnetic fields of approximately 30 mT were applied for 1 s with a quadrupole electromagnet integrated in the sample holder, along directions in 45° steps, so that equivalent states of D1 and D2 were sequentially prepared and captured (**Section 2.1**), and all images were acquired at remanence after the field had been removed.

*Registration and Magnetic Surface*: For the magnetic reconstruction, the signed-distance fields of D1 and D2 were averaged as described above. For each view, this geometry was placed in the frame of the respective device and projected onto the substrate plane along the X-ray direction. The projection was registered to the measured structural image $I^+$ by a two-dimensional similarity transformation with translation, in-plane rotation and a single isotropic scale factor, using the exterior outline and internal openings of the shadow, and the same transformation was applied to the corresponding XMCD images. The magnetic surface was defined as the continuous part of the upper surface whose local inclination, that is, the angle between the facet normal and the substrate normal, is below 75°. Because the sputtered flux arrives approximately perpendicular to the substrate, the film thickness deposited on a facet inclined by an angle $\theta$ scales approximately as $\cos\theta$.[14,46] Steeper facets receive less than about a quarter of the nominal thickness ($\cos 75° \approx 0.26$) and were not included. In the common device frame, the beam directions of the two views have azimuthal angles 45° apart and a shared elevation of $-16°$.

*Magnetization Reconstruction*: The coefficients $H_{vpf}$ in Equation 2 are the projected areas of the facet fragments crossed by the X-rays reaching pixel $p$, in units of the pixel area. Partial pixel coverage is retained, fragments of the same crossing are averaged with their projected areas, and distinct crossings add. Each facet magnetization is parametrized by one angle in its tangent plane. The loss function is $\mathcal{L} = E_d + \lambda_c E_c + \lambda_{\text{div}} E_{\text{div}}$, where $E_d$ is the mean squared difference between measured and predicted contrast, with both views weighted equally, $E_c$ penalizes differences between the magnetization of facets sharing an edge, and $E_{\text{div}}$ penalizes the discrete surface divergence of the magnetization. Both coefficients were set to $\lambda_c = \lambda_{\text{div}} = 1$. An approximate solution of the initial tangent-plane least-squares problem was computed with the LSMR iterative sparse least-squares solver,[53] and the inner optimization used the limited-memory Broyden-Fletcher-Goldfarb-Shanno algorithm with bound constraints (L-BFGS-B)[54,55], stopping when the relative change of $\mathcal{L}$ fell below $10^{-5}$. In order to reduce the influence of noisy data points on gain fitting, the gain factors were obtained by linear regression on single-pass pixels, excluding low-amplitude contrast comparable to the background noise (about 30% of the single-pass pixels). During magnetization refinement, the 5% of pixels with the largest absolute predicted contrast were excluded from the data-misfit term to limit the influence of extreme predicted values. This mask was refreshed after each gain update and held fixed during the subsequent inner optimization. The optimization was considered converged when the relative change of both gain factors remained below $10^{-4}$ for ten consecutive updates.
The first-phase outlier reset is an optional step used here to suppress isolated noise-driven reversals while the gain factors are still being refined. For each facet, the area-weighted mean magnetization is calculated within a lateral distance of approximately 80 nm along the connected magnetic surface. The mean vector is projected onto the facet's tangent plane and normalized, and replaces the facet magnetization when their angular difference exceeds 90°. This reset follows the fixed-gain field optimization and precedes the next gain update, reflecting the expectation that the remanent states prepared by a uniform field in a single direction do not contain isolated reverse domains on such a small scale. The reset is disabled in the second phase, allowing reverse magnetization supported by a sufficiently strong measured signal to be reconstructed through the balance among the data misfit, continuity penalty, and divergence penalty.

*Use of Generative AI*: The code for the geometry and magnetization reconstruction was developed with the assistance of the large language model ChatGPT (OpenAI, models 5.6 Solar and 6.1 Solar), used in 2026. The authors directed the development, reviewed, tested and revised all generated code, and take full responsibility for the code and the results.

**Acknowledgements**
This work was supported by the European Research Council (ERC) under the European Union's Horizon 2020 research and innovation program (grant agreement No. 101001290, 3DNANOMAG). This research was funded in whole or in part by the Austrian Science Fund (FWF) [10.55776/PIN1629824], through the TOMOCHIRAL project. For open access purposes, the author has applied a CC BY public copyright license to any author-accepted manuscript version arising from this submission. Funding from TU Wien's Blue Sky "Phoenix" program is acknowledged. A.A. acknowledges support through project 3D-Sky (Grant No. GAP-101108063), part of the Marie Skłodowska-Curie Actions under the European Union's Horizon Europe program. M.F. acknowledges funding from Agencia

Estatal de Investigación (AEI) through project PID2024-157112OB-C55. S.R.G. acknowledges funding from Agencia Estatal de Investigación (AEI) through project PID2024-155385NA-C32. We acknowledge the National Synchrotron Radiation Centre SOLARIS for providing access to the DEMETER beamline (proposal no. 251079), where the XMCD-PEEM measurements were performed. Research at the National Synchrotron Radiation Centre SOLARIS is supported by the Ministry of Science and Higher Education, Poland, under contract no. 1/SOL/2021/2.

**Conflict of Interest**

The authors declare no conflict of interest.

**Data Availability Statement**

The data that support the findings of this study will be openly available in TU Wien Research Data.

Nominally identical 3D nanostructures written at different azimuthal orientations replace mechanical sample rotation in shadow X-ray magnetic circular dichroism tomography. Using an indexable geometry reconstructed from electron microscopy silhouettes and a surface-tangent magnetization model, we combine projections from two differently oriented replicas to map the remanent magnetization of curved racetrack-like nanostructures after field initialization, revealing segments that switch and segments that persist.

**Orientation-Multiplexed Shadow XMCD-PEEM Tomography of 3D Nanomagnets**

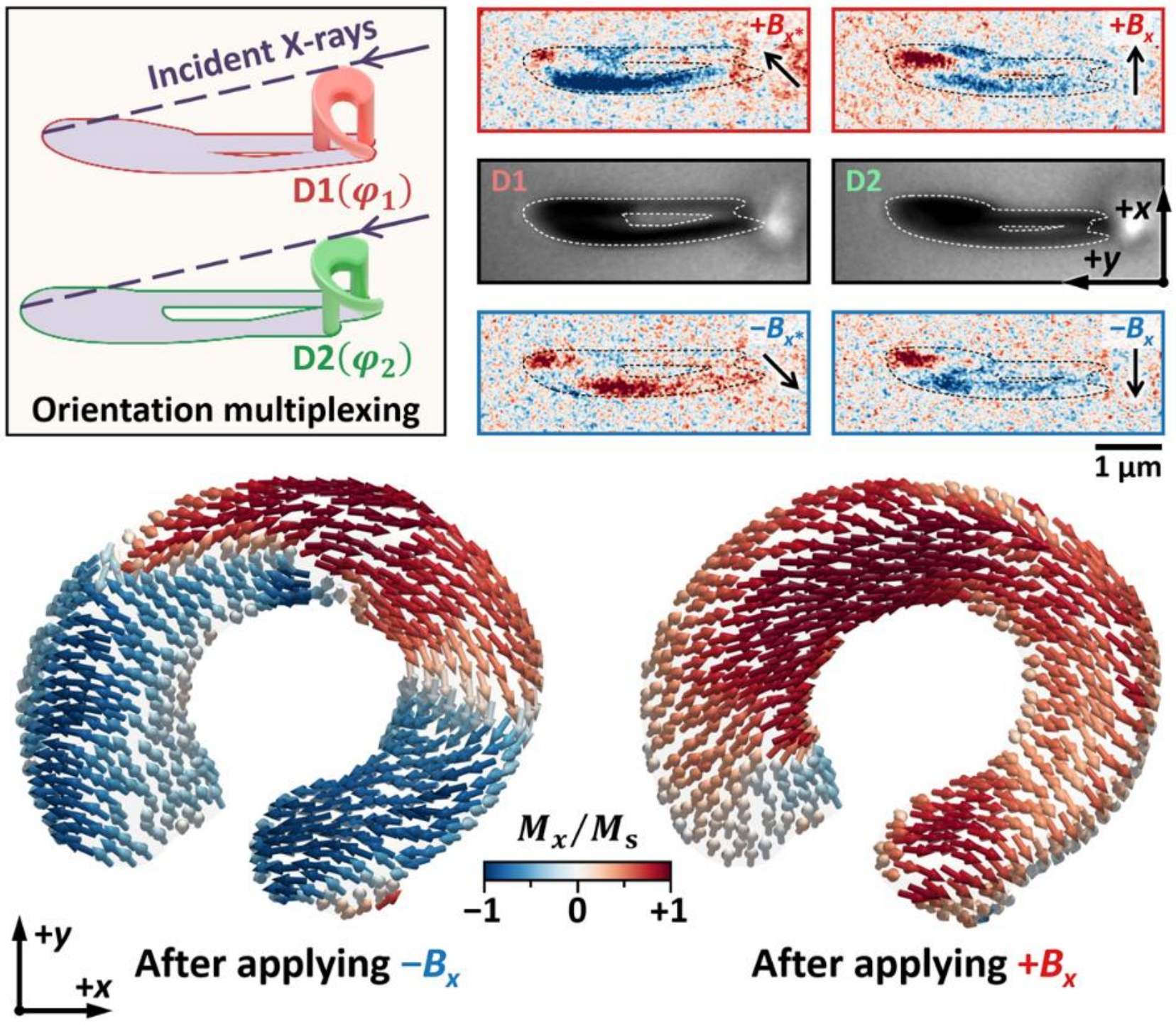